\documentstyle[preprint,aps]{revtex}

\begin{document}
\draft
\preprint{IMSc/2001/06/30}
\title{Isolated Horizon, Killing Horizon and Event Horizon}
\author{G. Date \footnote{e-mail: shyam@imsc.ernet.in}}
\address{The Institute of Mathematical Sciences,
CIT Campus, Chennai-600 113, INDIA.}
\maketitle
\begin{abstract} 
We consider space-times which in addition to admitting an isolated horizon 
also admit Killing horizons with or without an event horizon. We show
that an isolated horizon is a Killing horizon provided either (1) it
admits a stationary neighbourhood or (2) it admits a neighbourhood with
two independent, commuting Killing vectors. A Killing horizon is always
an isolated horizon. For the case when an event horizon is definable, all
conceivable relative locations of isolated horizon and event horizons are
possible. Corresponding conditions are given.
\end{abstract}

\vskip 0.50cm


\narrowtext

\section{Introduction}
Isolated horizons (IH) are defined quasi-locally and without any assumptions
about isometries (eg stationarity) of the space-time
\cite{isolated-1,isolated-2}. One also has completely quasi-local definitions
of mass, angular momentum etc such that the zeroth and the first law(s)
of black hole mechanics hold \cite{bh_laws,rotating}. Even quantum entropy
computations are available for the non-rotating isolated horizons
\cite{entropy}. These developments are non-trivial 
because firstly they are quasi-local and secondly they extend the scope
of black hole thermodynamics to space-times which need not be stationary.
This class of space-times is parameterized by infinitely many parameters
\cite{isometry}. Primarily it is the relaxation of condition of
stationarity that is the source of technical non-triviality, asymptotic
structure playing virtually no role. \\

IH therefore are candidates for replacing Killing horizons (KH) (for 
stationary black holes) as well as event horizons (EH) for purposes of black 
hole phenomena. Since all three horizons are distinct one could have space-times
wherein all are present. One could then ask if and how these are `related' to 
each other. How they are located relative to each other? Does existence of IH 
imply existence of the others when potentially possible? We address these
questions in stages.\\

We will consider space-times with an IH which could also admit Killing horizons.
Thus these space-times must have at least one isometry. We will show that
this alone is not sufficient to for an isolated horizon to be a Killing
horizon, further conditions are needed. However, every KH will be shown
to be an isolated horizon. Note that the question asked is different from asking
for the existence of any Killing vector at all \cite{isometry}. For this case 
asymptotic structure is irrelevant. \\

Next we will consider space-time which could have event horizon. We will
restrict to asymptotically flat (and strongly asymptotically predictable
\cite{wald-1}) space-times. We will be able to find conditions under
which an IH will imply existence of event horizon. For this case stationarity 
is not essential. The result is not immediately obvious since IH definition 
does not require foliation by marginally trapped surfaces.\\

The paper is organized as follows: In section II we recall the
definitions of IH, KH and EH and set the notation. In section III we
discuss the IH-KH relation while in section IV we discuss the IH-EH
connection. The last section contains summary and discussion. \\

The notation and conventions used are those of Chandrasekhar
\cite{chandra} with the metric signature (+ - - -)\footnote{
There is an inconsistent use of metric signature in Chandrasekhar's
book eg equations 287, 293 in the first chapter. In the definitions of
the Ricci scalars, equation 300, all the Ricci scalars and $\Lambda$
should be replaced by {\it{minus}} these scalars. Thus $\Phi_{00} ~=~ +
R_{11}/2. $ In reference \cite{isolated-2} this requires ${\cal{E}}$ to
be replaced by $-{\cal{E}}$ in equations 4 and 12. None of these affect
the conclusions of reference \cite{isolated-2}. The 18 equations of 
\cite{chandra} are also unaffected.}. The notation used
for isolated horizons is that of \cite{isolated-2}.%\\

\section{Preliminaries}

All space-times under consideration are solutions of Einstein-matter
equations with matter satisfying the dominant energy condition. \\

An isolated horizon $\Delta$, is defined as a null hypersurface with topology
$R\times S^2$, vanishing expansion of its null normal (non-expanding
horizon), and the Newman-Penrose spin coefficients $\mu, \lambda$ being
constant along the null generators. We also follow the procedure given
in reference \cite{isolated-2} of setting up the null tetrads on $\Delta$
and in an infinitesimal neighbourhood $U_{\Delta}$. Explicitly, the
following equations hold (modulo residual boost, constant scaling and
local rotation freedom for the choice of tetrads on $\Delta$):
\begin{center}
\begin{tabular}{lcl}
\mbox{In $U_{\Delta}$} ~&~:~&~$ \gamma ~=~ \nu ~=~ \tau ~=~ \mu -\bar{\mu} 
~=~ \pi - \alpha - \bar{\beta} ~=~ \kappa ~=~ 0.$ \\
\mbox{On $\Delta$} ~&~:~&~$ \rho ~=~ \sigma ~=~
{\underline{D}} \pi ~=~ 0 ~,~
\tilde{\kappa} ~\equiv~ \epsilon + \bar{\epsilon} ~~$ \mbox{(a non-zero
constant)} \\
\mbox{On $\Delta$} ~&~:~&~$
\Psi_0 ~=~ \Psi_1 ~=~ \Phi_{00} ~=~ \Phi_{01} ~=~ 0 .$ \mbox{(energy
conditions, Raychoudhuri equation)}\\
\mbox{On $\Delta$} ~&~:~&~$ {\underline{D}} \lambda ~=~ {\underline{D}} \mu 
~=~ 0 ~~$ \mbox{(Isolated horizon conditions)}.
\end{tabular}
\end{center}

The underlined derivatives are the rotation covariant derivatives
(`compacted' derivatives) \cite{isolated-2}. \\

Note that a non-expanding horizon admits several equivalence classes of
null normals, $[\ell]$ while isolated horizon conditions select a unique
equivalence class. The choice of tetrads, conditions on spin
coefficients etc are available for every equivalence class on a
non-expanding horizon.\\

A Killing horizon is defined as \cite{wald-2} a null hypersurface, again
denoted by $\Delta$, such that a Killing vector is normal to $\Delta$. This 
clearly needs existence of a Killing vector at least in a neighbourhood of 
$\Delta$. Note that Killing vector need not exist everywhere in the
space-time. It is also not required, a priori, to be time-like. As such the
topology of $\Delta$ is not stipulated. For the questions explored here
however we will consider only those Killing horizons which are
topologically $R \times S^2$. This will be important in the following. \\

Definition of an event horizon needs a notion of `infinity' and its `structure'
to be specified. We will take this to be asymptotically flat. We will
also take the space-time in this context to be strongly asymptotically
predictable \cite{wald-1} in order to admit possibility of the usual
black holes (not necessarily stationary). If the causal past of the
future null infinity does not equal to the space-time, one says that a
black hole region exists. Its boundary is then defined to be the event
horizon. In the asymptotically flat and stationary context, the topology of 
event horizon is $R \times S^2$. In the following, we will restrict to 
asymptotically flat context and assume that event horizons under consideration 
are all smooth and with topology $R \times S^2$. 

\section{Isolated Horizon and Killing Horizon}

A killing horizon is automatically a non-expanding horizon since the
induced metric on the leaves is preserved by the diffeomorphism
generated by the Killing vector and hence so are the areas of the
leaves. Also, an isolated horizon is a non-expanding horizon with
further conditions. Consider therefore a solution admitting a
non-expanding horizon, $\Delta$ with topology $R \times S^2$ and an 
infinitesimal neighbourhood $U_{\Delta}$. We also assume that $\Delta$
is also ``isolated" in the sense that in $U_{\Delta}$ there is no other
non-expanding horizon. This will ensure that when a Killing vector exist
in $U_{\Delta}$, it will be necessarily tangential to $\Delta$
\footnote{I thank Abhay Ashtekar for alerting me to the need for this
assumption.}. All the machinery of null tetrads etc described in the previous 
section is available. We are interested in finding out if a Killing horizon 
is an isolated horizon and conversely. To this end assume further that the 
solution also admits at least one Killing vector, at least in $U_{\Delta}$. \\

We want to see that if the Killing vector is normal to $\Delta$, does
there exist an equivalence class of null normals such that $\Delta$
becomes an isolated horizon. For the converse we first note that an 
isolated horizon does {\it{not}} imply existence of an isometry even in 
a neighbourhood $U_{\Delta}$ \cite{isometry}. However we are given that 
an isometry exists and the question is whether there exists {\it{a}} Killing 
vector {\it{normal}} to the IH. \\

To this end let us write a Killing vector in the form (in $U_{\Delta}$):
\begin{equation}
\xi^{\mu} ~\equiv~ A~\ell^{\mu} ~+~ B~n^{\mu} ~+~ C~m^{\mu} ~+~
\bar{C}~\bar{m}^{\mu}
\end{equation}
The Killing equations, valid in $U_{\Delta}$, can be written in terms of the 
$A, B, C, \bar{C}$ as:
\begin{eqnarray}
D^{\prime} A ~& = &~ 0 ~;\\
D^{\prime} B ~& = &~ -D A - \tilde{\kappa} A -
\bar{\pi} C - \pi \bar{C}~; \\
D^{\prime} C ~& = &~ \bar{\delta} A + \pi A +
\mu C + \lambda \bar{C}; \\
D B ~& = &~ \tilde{\kappa} B; \\
D C ~& = &~ - (\epsilon - \bar{\epsilon}) C + \bar{\delta}
B - 2 \pi B - \rho C - \bar{\sigma} \bar{C}; \\
\bar{\delta} C ~& = &~ - (\alpha - \bar{\beta}) C + \bar{\sigma} A - \lambda B
\\
\delta C ~& = &~ (\bar{\alpha} - \beta) C - \bar{\delta} \bar{C} - (\alpha -
\bar{\beta}) \bar{C} + (\rho + \bar{\rho}) A - 2 \mu B
\end{eqnarray}

{\underline{Remarks:} \\

(1) These equations can be thought of `evolution' equations for $A, B,
C, \bar{C}$ to go off-$\Delta$ with their values being specified on $\Delta$.
Thus for any consistent choice of $A, B, C$ on $\Delta$, we can obtain Killing 
vector in $U_{\Delta}$. On the horizon, the Killing vector must be tangential 
to the horizon and therefore, $B = 0$ on $\Delta$. The special case
where $B = C = 0, A \ne 0$ on $\Delta$ corresponds to the non-expanding
$\Delta$ being a Killing horizon. \\

(2) On $\Delta$, $C$ satisfies, ${\underline{D}}C = 0 =
{\underline{\delta}}C + \bar{\underline{\delta}} \bar{C} $.
Derivatives of $A$ along $\Delta$ and one combination of derivatives of $C$ 
along leaves is not explicitly specified. \\

The coefficient functions however must satisfy the commutator identities 
\cite{chandra,isolated-2}. These applied to $C$ and $B$ imply, on $\Delta$ : 

\begin{eqnarray}
{\underline{\bar{\delta}}}^2 A ~& = &~ - 2 \pi
\underline{\bar{\delta}} A + \lambda ( D A + 2 \underline{\delta} C )
- (C \underline{\delta} + \bar{C} \underline{\bar{\delta}}) \lambda 
- A \underline{D}\lambda \\
\underline{\delta} ~ \underline{\bar{\delta}} A ~& = &~ - (\pi 
\underline{\delta} + \bar{\pi} \underline{\bar{\delta}}) A + \mu D A -
(C \underline{\delta} + \bar{C} \underline{\bar{\delta}}) \mu 
- A \underline{D}\mu \\
\underline{\bar{\delta}} ~ \underline{D} A ~& = &~ - \tilde{\kappa} 
\bar{\underline{\delta}} A -
(C \underline{\delta} + \bar{C} \underline{\bar{\delta}}) \pi + \pi 
\underline{\delta} C \\
\underline{D}^2 A ~& = &~ - \tilde{\kappa} \underline{D} A 
\end{eqnarray}

We have deliberately {\it{not}} used the isolated horizon conditions.\\

Now if $\Delta$ is a Killing horizon for the above Killing vector, then 
$A \ne 0$ and $C = 0$ on $\Delta$. It remains to check if the isolated 
horizon conditions hold. \\

For the special case of $B = C = 0$ on $\Delta$, the equations
simplify. The equations (11) and (12) imply that ${\underline{D}}A +
\tilde{\kappa} A = Q$ where $Q$ is a constant. It is easy to see that
$Q$ is the acceleration of the Killing vector (on $\Delta$). The off-$\Delta$
derivative of the norm of the Killing vector, evaluated on the horizon is 
$-2 A Q$ while the norm itself is of course zero. \\

%Thus if either of $A,
%Q$ is zero, the Killing vector is null to first order, off-$\Delta$. We
%assume this to be not the case, i.e. neither $A$ nor $Q$ is zero on the
%horizon. This immediately implies that the Killing vector is
%{\it{time-like}} at least on `one side' of $\Delta$.\\

A Killing vector can not vanish on a hypersurface without vanishing
everywhere \footnote{I thank Abhay Ashtekar for drawing my attention to
this fact.} and thus $A \ne 0$ on $\Delta$. 
Now by making a scaling transformation $\ell \rightarrow A^{-1} \ell$,
we go to another equivalence class of null normals in which the Killing
vector is just the new null normal, i.e. effectively, $A = 1$. This
immediately gives $Q = \tilde{\kappa}$. Furthermore, equations (9) and
(10) imply the isolated horizon conditions! For the case of non-zero
surface gravity, the off-$\Delta$ derivative of the norm of the Killing
vector is then non-zero and hence the Killing vector is {\it{time-like}}
at least on `one side' of $\Delta$.\\

Thus we have shown that {\it{if a non-expanding horizon is a Killing horizon,
then it admits an equivalence class $[\ell]$ with respect to which it is
an isolated horizon.}}\\

To consider the converse we now take $\Delta$ as an isolated horizon.\\

Now we have two possibilities. Either $C = 0$ on $\Delta$ or it is
non-zero. In the former case the Killing vector which is given to exist, 
is normal to the horizon and hence $\Delta$ is a Killing horizon. As
seen above this means that the Killing vector is time-like on `one side'
of $\Delta$ at least when the surface gravity is non-zero. Conversely, if the 
Killing vector is time-like at least on `one side' of $\Delta$ (i.e. $AB > 
C\bar{C}$ ), then on $\Delta$, it must be normal to $\Delta$ making it a 
Killing horizon. \\

In the latter case the norm of the Killing vector on $\Delta$ is negative.
In this case, the $D^{\prime}$ of the norm of the Killing vector being non-zero
on $\Delta$ and norm itself being negative on $\Delta$ implies that the
Killing vector is space-like in $U_{\Delta}$. This is for instance
realized by a Killing vector of `axisymmetry'. Clearly, mere existence of
a Killing vector in an neighbourhood of an isolated horizon is
{\it{not}} enough for the IH to be a Killing horizon. Norm being
positive on `one side' is needed in addition. This could be ensured by
existence of two independent commuting Killing vectors in a
neighbourhood. This is seen as follows. \\

Suppose that $\xi_1, \xi_2$ are two Killing vectors in $U_{\Delta}$
which commute. This implies that there are two independent, mutually commuting
isometries of $\Delta$ (the horizon is a hypersurface, so the
independence of Killing vectors in the neighbourhood continues to hold
on the horizon). By repeating the arguments of reference
\cite{isolated-2} one can choose a unique foliation such that one of the
Killing vector, $\xi_1$, is tangential to the leaves ($A_1 = 0$). The
commutativity of the Killing vectors and the spherical topology of
leaves imply that the second Killing vector can {\it{not}} be tangential to the
leaves i.e. $A_2 \ne 0$. Now however we can construct a linear
combination, $\xi_3$, of these Killing vectors which has $A_3 \ne 0, C_3
= 0$. Commutativity of the Killing vectors is used for this statement. This 
Killing vector is normal to $\Delta$ implying that $\Delta$ is
a Killing horizon. The linear combination is the precise analogue of the
combination appearing in the Kerr solution. The commutativity of the
Killing vectors and the spherical topology of the leaves is essential for
this argument. \\

Thus we conclude that for an isolated horizon to be a Killing horizon
there must exist a neighbourhood together with {\it{either}} one Killing vector 
which is time-like at least on `one side' of $\Delta$ {\it{ or}} two commuting 
Killing vectors.  In the former case the Killing vector is trivially
normal to $\Delta$ while in the latter there exist another Killing
vector which is normal to $\Delta$. The two cases of course correspond to
the usual static and the stationary-axisymmetric examples. Completeness
of Killing vectors (group of isometries) is not essential for this argument.\\

It would be nicer to have a condition on isolated horizon spin
coefficients directly which will guarantee either of the above
conditions. One could of course try to {\it{define}} a Killing vector by
taking $A = $ constant, $B = C = 0$ on $\Delta$. The argument of Lewandowski 
\cite{isometry} shows that for the special case of non-rotating,
spherically symmetric, vacuum isolated horizon such a definition is in 
conflict with non-zero $\Psi_4$. However for general isolated horizons it 
is not clear what `obstructions' could be there. Except for the special case
mentioned above or the case where the leaves have at least one isometry
of their intrinsic metric, the foliations are not unique and 
$\Psi_3, \Psi_4$ are not invariant under residual boosts. \\

\section{Isolated Horizon in potentially black hole space-times}

To admit the possibility of a (usual) black hole we will consider space-times
which are strongly asymptotically predictable and are solutions of
Einstein-matter equations with matter stress tensor satisfying the
dominant energy condition. In addition, we assume that such a space-time
also has an isolated horizon. Note that there is however {\it{no}}
condition on $\mu$ being positive/negative on $\Delta$ nor is there any
assumption made about the symmetry class of the horizon. \\

We could of course have a special case where $\Delta$ admits a foliation
such that $\mu < 0$ on a leaf and hence on $\Delta$. In this case the
horizon is foliated by marginally trapped surfaces \cite{wald-1} and hence
belong to the black hole region, $M - J^-(\cal{J}^+)$, of the
space-time i.e. $S^2 \cap J^-(\cal{J}^+)$ = $\emptyset $. In particular, 
an event horizon (EH) must exist. \\

Now the IH and EH may or may not intersect. If they do not intersect,
then the IH is irrelevant for any observer outside the event horizon. If
however they do intersect then they must do so on a space-like two
dimensional surface, in fact $S^2$ or IH must be a subset of EH. In the
former case null generators of the EH will be either along $\ell$ or along 
$n$ elements of the tetrad on IH. But the expansion of the generators of EH 
is non-negative (area theorem) while $\mu$ is negative. Hence the generators 
of EH and IH must be proportional at points of intersections. The isolated 
horizon being non-expanding now implies that the expansion of the generators 
of the EH must also be zero and {\it{it can not become positive in the future}}.
Thus the area of the instantaneous black hole corresponding to the
coinciding portion of the event horizon must remain constant until
possible future merger with other instantaneous black holes. Thus, 
in the special case of $\mu < 0$, we see that (a) event horizon must exist  
and (b) either IH is irrelevant or it coincides with an instantaneous
black hole (a connected component of intersection of EH with a Cauchy
surface). Its area can change only with possible future mergers with
other black holes. Thus such an isolated horizon refers to the settled 
stage of an instantaneous black hole. \\

%% Really the area elements, we need to check if IH intersection EH is
%% S^2

However a general isolated horizon does not require (or imply) existence
of a foliation such that $\mu < 0$ on a leaf. The same proof which shows
that marginally trapped surfaces are invisible from future null infinity
can be used to explore sub-cases of the general case. The argument is as
follows. \\

Let $T$ be a leaf in a foliation of the isolated horizon. It is a
compact, orientable, space-like submanifold of the space-time. Hence the
boundary of its causal future is generated by null geodesics starting
orthogonally from $T$ and having no conjugate points (theorem 9.3.11 of
\cite{wald-1}). Consider now $X \equiv T \cap J^-(\cal{J}^+)$. If $X$ is
empty, then IH again implies existence of EH and is contained in the
black hole region. Once again IH and EH may
not intersect making the IH irrelevant or IH may coincide with a portion
of the EH or they may intersect non-trivially in a space-like $S^2$. If
however $X$ is non-empty, then boundary of causal future of $T$ contains
a whole spherical cross-section of ${\cal{J}}^+$. In the vicinity of
future null infinity, the expansion of any null geodesic congruence
orthogonal to a cross-section of the null infinity is strictly positive.
Hence on a spherical cross section on the boundary, the generators must
have positive expansion. Since these generators can not have conjugate
points, we must have their expansions on $T$ itself to be strictly
positive. These generators thus must comprise of the $n$ congruence and
hence $\mu > 0$ must hold on $T$. The condition of isolation then
implies that $\mu$ is positive on $\Delta$. \\

The Gauss-Bonnet integral \cite{isolated-2} however is:

\begin{equation}
\tilde{\kappa} ~ \int_{\Sigma_2} \mu ~ = ~ - 2 \pi + 
\int_{\Sigma_2} \left( {\cal{E}} + \pi\bar{\pi} \right)
\end{equation}

From this we see that if the surface gravity is positive then we must
have the integral on the right hand side to be larger than $2 \pi$. If the 
IH is non-rotating then we can choose a foliation such that $\pi = 0$  and 
if in addition it is a vacuum solution then $\mu$ can not be strictly positive 
on $\Delta$. But this means that $X$ must be empty and hence an EH must exist. 
Thus a non-rotating isolated horizon with positive surface gravity in a 
space-time which is vacuum on the horizon, necessarily implies existence of 
an event horizon. In particular this implies that in {\it{Minkowski space-time,
there can not be a non-rotating isolated horizon with topology $R \times S^2$}}.
However, for a rotating isolated horizon and/or space-time which is non-vacuum 
on the IH, it is possible that the isolated horizon is visible from future null 
infinity {\it{provided the Gauss-Bonnet integral so permits}}. Note that in 
this case an event horizon could well exist but is not implied by the existence 
of an IH. \\

Since now we could have an IH outside of an EH for the rotating and/or
non-vacuum case the IH itself is accessible from infinity unlike an EH. \\
%This 
%is important in that an ``observable" black hole may now be identified with 
%the presence of an isolated horizon. This analysis is independent of any
%isometries.\\

\section{Summary and discussion}

In this work we have explored the implication of existence of an isolated 
horizon (topologically $R \times S^2$) for the existence of an event horizon 
when such is applicable and for Killing horizons when such are
admissible. \\

For the relationship between a Killing horizon and an isolated horizon
we have shown that a Killing horizon is always an isolated horizon but
for the converse we need either a Killing vector which is time like on
one side of the IH or we need two commuting Killing vectors. Mere
isometry of space-time together with an isolated horizon does {\it{not}}
imply existence of a Killing horizon, one needs either `staticity' or
additional `axisymmetry'. \\

For the relationship between IH and EH, we have shown that for the 
non-rotating isolated horizon in a solution (vacuum near $\Delta$) an event 
horizon must necessarily exist. Furthermore it corresponds (when
relevant) to the settled stage of a black hole (no further area increase 
unless merger with other instantaneous black holes occurs). For the case of 
a rotating isolated horizon with or without non-trivial ${\cal{E}}$, we also 
have the possibility of no event horizon or at least accessibility of the IH 
from future null infinity.\\

Consider the case where we have several isolated horizons all coinciding
with portions of the event horizon. Then the area theorem for event
horizon implies that the isolated horizons which are to the future of a
given isolated horizon will have areas larger than or equal to the area
of the given IH. In this sense a second law of black hole mechanics can
be seen to generalize to a family of isolated horizons. \\

%The possibility that 
%one could have an isolated horizon out side the event horizon means that a 
%different argument is be needed to interpret a second law for isolated 
%horizons. \\

The initial restriction to stationary black holes was justified on the
grounds of uniqueness theorems and that a stationary black hole, if
disturbed, will return quickly to another stationary black hole. For
isolated horizons, no such results are available. Since an isolated
horizon need not imply an event horizon or even when an EH is present,
an IH could be outside of it means that the settling down of a disturbed
IH is a non-trivial issue and so is a physical process interpretation
of the first law.\\

{\underline{Acknowledgments:}} The author gratefully acknowledges
useful discussions with Dr. Sukratu Barve during the initial stages of
this work. It is a pleasure to thank Abhay Ashtekar for a critical
discussion during the GR16 meeting which helped tighten some of the
arguments.

\end{document}